\documentclass[english,aps,reprint,onecolumn,amsfonts,amsmath,amssymb,superscriptaddress]{revtex4-1}

\usepackage[T1]{fontenc}
\usepackage[utf8]{inputenc}
\usepackage{hyperref}
\usepackage{graphicx}
\usepackage{babel}
\usepackage{subfigure}
\usepackage{color}
\usepackage{amsmath}
\usepackage{changepage}

\begin{document}

\title{Unsupervised noise reductions for gravitational reference sensors or accelerometers based on Noise2Noise method}

\author{Zhilan Yang}
\affiliation{Taiji Laboratory for Gravitational Wave Universe (Beijing/Hangzhou), University of Chinese Academy of Sciences, Beijing 100049, China.}
\affiliation{Center for Gravitational Wave Experiment, National Microgravity Laboratory, Institute of Mechanics, Chinese Academy of Sciences, Beijing 100190, China.}

\author{Haoyue Zhang}
\affiliation{Lanzhou Center of Theoretical Physics, Lanzhou University, Lanzhou 730000, China.}
\affiliation{Center for Gravitational Wave Experiment, National Microgravity Laboratory, Institute of Mechanics, Chinese Academy of Sciences, Beijing 100190, China.}

\author{Peng Xu}
\email{xupeng@imech.ac.cn}
\affiliation{ICenter for Gravitational Wave Experiment, National Microgravity Laboratory, Institute of Mechanics, Chinese Academy of Sciences, Beijing 100190, China.}
\affiliation{Taiji Laboratory for Gravitational Wave Universe (Beijing/Hangzhou), University of Chinese Academy of Sciences, Beijing 100049, China.}
\affiliation{Laboratory of Gravitational Wave Precision Measurement of Zhejiang Province, Hangzhou Institute for Advanced Study, UCAS, Hangzhou 310024, China.}
\affiliation{Lanzhou Center of Theoretical Physics, Lanzhou University, Lanzhou 730000, China.}

\author{Ziren Luo}
\affiliation{ICenter for Gravitational Wave Experiment, National Microgravity Laboratory, Institute of Mechanics, Chinese Academy of Sciences, Beijing 100190, China.}
\affiliation{Taiji Laboratory for Gravitational Wave Universe (Beijing/Hangzhou), University of Chinese Academy of Sciences, Beijing 100049, China.}
\affiliation{Laboratory of Gravitational Wave Precision Measurement of Zhejiang Province, Hangzhou Institute for Advanced Study, UCAS, Hangzhou 310024, China.}

\begin{abstract}
    Onboard electrostatic suspension inertial sensors are important applications for gravity satellites and space gravitational wave detection missions, and it is important to suppress noise in the measurement signal. Due to the complex coupling between the working space environment and the satellite platform, the process of noise generation is extremely complex, and traditional noise modeling and subtraction methods have certain limitations. With the development of deep learning, applying it to high-precision inertial sensors to improve the signal-to-noise ratio is a practically meaningful task. Since there is a single noise sample and unknown true value in the measured data in orbit, odd-even sub-samplers and periodic sub-samplers are designed to process general signals and periodic signals, and adds reconstruction layers consisting of fully connected layers to the model. Experimental analysis and comparison are conducted based on simulation data, GRACE-FO acceleration data and Taiji-1 acceleration data. The results show that the deep learning method is superior to traditional data smoothing processing solutions. 

\end{abstract}

\maketitle

\section{Introduction \label{Intro}}

Inertial sensors are able to detect acceleration, angular velocity, gravity vectors, and other motion factors. This paper focuses on the application of inertial sensors in the fields of aviation and gravitational wave detection.
In space, star trackers can serve as the inertial reference for spacecraft, and are employed in space satellite missions related to precise space measurements, high-precision navigation, and mapping. High-precision inertial sensors play a crucial role as key scientific payloads in space missions such as global gravity field inversion, space gravitational wave detection, and gravitational field measurements. Among modern advanced technologies including superconducting magnetic levitation and SQUID technology, and cold atom interferometry technology, high-precision electrostatic suspension inertial sensors based on electrostatic servo control technology are still the most advanced and reliable onboard inertial reference technology for in-orbit operation. Accomplishing a series of gravity satellite missions, including GRACE \cite{tapley2004grace}, GRACE-FO \cite{bandikova2019grace}, and GOCE \cite{drinkwater2003goce}, high-precision inertial sensors that use electrostatic levitation have proven to be successful. The ultra-high-precision electrostatic suspension inertial sensor, verified by the LISA PathFinder \cite{armano2015lisa} satellite, is expected to become a crucial payload for upcoming space gravitational wave detection missions, such as the LISA \cite{armano2015lisa}, Taiji \cite{luo2020brief}, and Tianqin plans \cite{luo2016tianqin}
 In the Taiji mission, inertial sensors can not only track the motion of the test mass by measuring the capacitance change caused by its displacement but also measure the acceleration signal of the test mass with high accuracy. The accuracy of the inertial sensor has an important impact on the sensitivity of space gravitational wave detection missions.The electrostatic suspension inertial sensor system mainly consists of the test mass as the inertial reference, the electrode housing with sensing and driving electrodes, the servo control electronic system, the vacuum system, and other charge management subsystems and isolation subsystems that depend on the specific space environment and satellite platform characteristics for a given mission. From a hardware perspective, it is challenging to minimize the effects of environmental interference and noise on the measurement system and achieve higher detection sensitivity through hardware improvements once a certain level is reached.
Due to its extremely high sensitivity, the on-board electrostatically suspended inertial sensor is subject to complex physical environment coupling interference, including temperature gradient fluctuations, magnetic fields, electric field fluctuations, microvibrations, high-energy particle beam charging, and other environmental factors related to the specific space environment and satellite platform characteristics. Moreover, the measurement data of the sensor is extremely complex with interference and noise components, including readout noise and control noise. Therefore, traditional processing methods such as noise modeling, subtraction, data smoothing, and trend fitting are used to suppress noise and improve signal-to-noise ratio. However, the noise singal often exhibits unfavorable fluctuations in frequency similar to potential scientific signals, making it difficult to achieve the best noise filter \cite{morvan2022don}. Intelligent computing methods are considered to assist in completing this task. This is a critical technology for further improving the detection sensitivity of inertial sensors and achieving accurate and efficient scientific applications of measurement data.

In recent years, deep learning has developed rapidly and made progress in many fields. For example, in the fields of Computer Vision, Natural Language Processing and Time-series, deep learning methods continue to evolve and reach advanced levels of noise reduction compared with traditional processing methods. Generally speaking, deep learning techniques can be divided into two categories: supervised and unsupervised learning. With supervised methods, we need ture data to train the neural network model. Changhui Jiang et al. \cite{jiang2018mems},based on supervised learning, used a combination of recursive neural network (RNN) and long and short time memory network (LSTM) to process the output of inertial sensor as a time series signal, finally improving the accuracy of inertial sensor. Despite the wide application of supervised learning, its limitations are evident, as it is difficult to acquire data that is close to true values and the accuracy of the true value samples severely affects the effectiveness of denoising algorithms. In the case of signal measurement, such as signals from satellites like Taiji-1 and GRACE-FO, only noisy signals but no true signals are obtained. Although it is possible to obtain data from simulating dynamic equations constructed by sensors and observed objects, or from ground tests simulating orbit environmental conditions, the data obtained by these methods are often not accurate enough, thus affecting the final denoising effects. Therefore, considering unsupervised learning methods for signal denoising is recommended. For unsupervised or self-supervised learning denoising, Jaakko Lehtinen et al. \cite{2018Noise2Noise} proposed the Noise2Noise(N2N) denoising framework, which can train a denoising network model without using clean images as training samples. The authors experimented with simple noise distributions (Gaussian, Poisson, Bernoulli) and complex, intractable synthetic noise from Montalcaro images. The final results show that deep learning neural networks are able to denoise signals without the need for clean target data, and the final performance level is equal to or close to that of using clean target data. In order to solve the limitation of requiring multiple noise samples in the N2N method, Tao Huang et al. \cite{huang2021neighbor2neighbor} proposed the Neighbor2Neighbor method based on the N2N method, and proposed a random neighbor subsampler to generate training image pairs based on a single noise image sample. The final experimental results show that the deep learning method can deal with the problem of only single noisy signal denoising, and has achieved higher effectiveness and superiority than the existing methods. In addition to the field of image, many people have applied the N2N method to the signal processing of time series. Qingchun Li et al. \cite{li2021speech} proposed a Single noise audio denoising framework (SNA-DF) based on N2N for processing single noise audio denoising, and used the deep and complex U-net model to realize the denoising processing. Shirong Koh et al. \cite{koh2020underwater} proposed the WaveN2N model to deal with the noise of acoustic signals in underwater areas without the prior knowledge and clean signals from real data; Takayuki Takaai et al. \cite{takaai2021unsupervised} applied the Noise2Noise method to current waveform signals obtained from multi-stage narrow nanochannels, which are characterized by high noise and complex measurement principles. The CAE model and U-net model are used respectively, and the final noise reduction effect is better than the traditional signal processing methods such as frequency filter, wavelet transform and Kalman filter, which can retain the signal details more accurately; Noise reduction framework is not only limited to Noise2Noise. Mario Morvan et al.  \cite{morvan2022don} 's Transformer model based on Noise2Self framework combined with time series realizes light curve signal noise reduction for damaged TESS models. This method has flexibility and better performance when dealing with large data sets. The main contributions of this work are as follows:
\begin{enumerate}
\item	We applied the N2N method for the first time to suppress noise in inertial sensor data.
\item   The signal was divided into periodic and general components, and we proposed the use of a periodic sub-sampler and odd-even sub-sampler. For the periodic component, we proposed the addition of a reconstruction layer to the model.
\item   We applied the N2N method to the Taiji-1 calibration task and GRACE-FO Level-1A data, effectively suppressing noise.
\end{enumerate}

The structure of this paper is as follows:Section \uppercase\expandafter{\romannumeral2} introduces the working principle and noise analysis of electrostatic suspension inertial sensor. Section \uppercase\expandafter{\romannumeral3} introduces an overall denoising framework, the theoretical basis of N2N method, and the custom neural network architecture, Section \uppercase\expandafter{\romannumeral4} introduces the simulation experiment and real experiment about Taiji-1 satellite and GRACE-FO satellite, and also carries out the comparative research results and discussion with a variety of filters. Finally, Section \uppercase\expandafter{\romannumeral5} summarizes the results of the work as well as the potential range of applications.

\section{Electrostatic Levitation Inertial Sensors\label{IS}}
\subsection{overview}

The electrostatic levitation inertial sensor system comprises several primary subsystems, including the test mass, electrode housing that surrounds the test mass, electronic measurement and control unit, and vacuum chamber, with the test mass serving as the fundamental inertial reference component. Depending on the mission environment and satellite platform characteristics, supplementary subsystems such as charge management and system isolation may also be incorporated.
The data experiments conducted in this paper are based on available GRACE-FO inertial sensor (or accelerometer) Level-1A data and measured data from the Taiji-1 satellite inertial sensor. The basic principles of the inertial sensors used in both missions are similar. The distribution of the test mass and its surrounding measurement and control electrodes in the inertial sensor are illustrated in Figure 1.

\begin{figure}[htbp]
\includegraphics[width=10.5 cm]{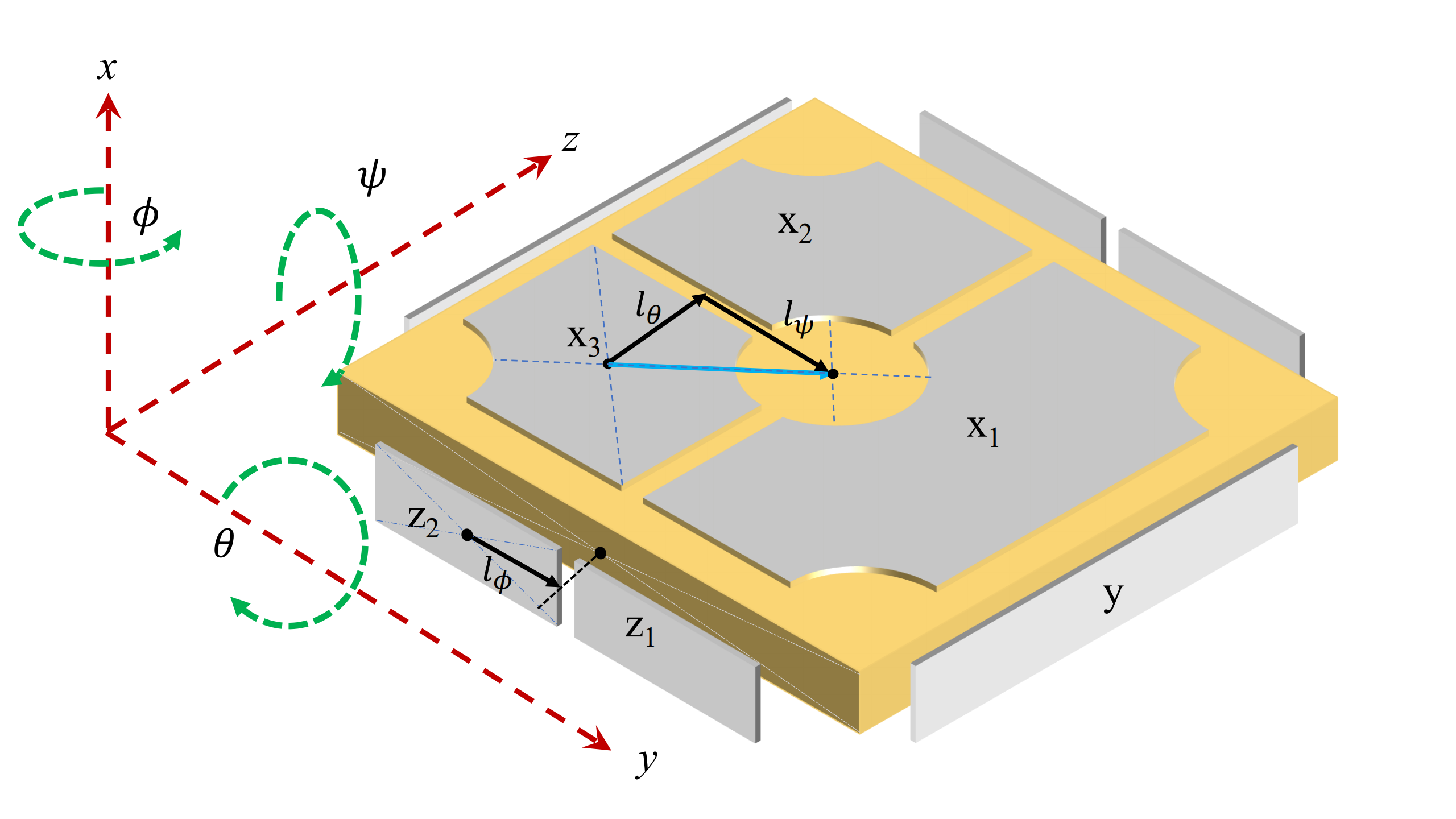}
\caption{Arrangement of the primary mechanical components of the inertial sensor.}
\end{figure}

The test mass is located at the nominal geometric center position inside the electrode housing. The sensitive structure consisting of the test mass and the electrode housing is situated in a vacuum environment with stable magnetic field fluctuations and temperature gradients. During scientific operations of the spacecraft, deviations from the inertial trajectory of free fall in the gravitational field occur due to non-gravitational disturbances from the space environment, resulting in relative motion between the satellite platform and the test mass. The resulting relative displacement is projected onto a measurement axis, causing changes in the distance between the test mass surface and the corresponding electrodes on either side, which is reflected in changes in capacitance. The differential capacitance detection circuit precisely measures the displacement of the test mass relative to its nominal center position. The control unit receives this data and adjusts the electrostatic force by changing the voltage, in order to real-time control the relative motion between the test mass and the electrode housing, and maintain the test mass in the vicinity of the nominal equilibrium position.
\subsection{the working principle of inertial sensors}
In actual operation, electrostatically suspended inertial sensors on-board spacecraft have two operating modes, namely, the accelerometer mode and the drag-free control mode. In missions such as LISA PathFinder and Taiji-1, the inertial sensors on the space gravitational wave detection technology experimental satellite can switch between these two modes depending on the mission requirements. For gravity satellite missions such as GRACE and GRACE-FO, the inertial sensors operate in the accelerometer mode.
In the accelerometer mode, the electrostatic force is applied to the test mass by changing the voltage on the electrodes surrounding the test mass to keep it in its nominal equilibrium position. In the drag-free mode, the object of control is changed, and the satellite is controlled by feedback control applied to the thrusters, so that the satellite follows the motion of the test mass, which is suspended near its equilibrium position. Taking the accelerometer measurement mode as an example, the residual relative acceleration is negligible.The resulting non-gravitational perturbations to the spacecraft (atmospheric drag, solar pressure, Earth reflection, etc.) are as follows:
\begin{equation}
a_{para, SC }^i(t)=-p_1^{i \alpha} V_\alpha(t)-G^{i j}(t) d_j-a_{para, TM}^i(t)-k^i
\end{equation}
where $a_{TM}^i(t)$ and $a_{SC}^i(t)$ respectively denote the accelerations of the test mass and spacecraft relative to the inertial reference frame, $k^i$ represents the acceleration bias and $p^{i\alpha}$ represents the linear acceleration scale factor. The subscript "para" signify the parasitic disturbance acceleration. ${d}$ is the deviation between the test mass centroid and the spacecraft centroid \cite{zhang2023systematic}.

After the center of mass deviation has been corrected and working parameters have been accurately calibrated, control voltage or control acceleration can be used to provide precise measurements of non-gravitational disturbances acting on the spacecraft. The level of parasitic acceleration disturbance to the test quality in the equation is a key factor that limits measurement accuracy.

\subsection{noise analysis of accelerometers}
During the actual operation of the satellite-borne electrostatic inertial sensors, due to their extremely high sensitivity and the complex coupling relationship between the test mass and the surrounding physical fields of multiple physical fields, the components of the parasitic disturbance acceleration $a^i_{para,TM}(t)$ that the test mass in Equation (1) experiences are extremely complex. These mainly include: acceleration disturbances caused by the measurement and control of the test mass, including displacement detection noise and control errors caused by it, and acceleration noise caused by unstable control voltage; acceleration noise caused by the coupling of the relative displacement jitter and parasitic stiffness (mainly electrostatic stiffness and self-gravity stiffness) of the test mass with respect to the equilibrium position; acceleration noise caused by the coupling of the residual magnetic moment and induced magnetic moment of the test mass with magnetic field fluctuations; noise generated by residual gas molecule random collisions, residual gas molecule coupling temperature gradient fluctuations of thermal radiometer effect in vacuum environment, as well as asymmetric gas outflow effect; acceleration noise generated by the coupling of test mass charge fluctuations with electric field environment; and acceleration noise generated by dissipative effects including dielectric loss.

The majority of the aforementioned noises cannot be precisely eliminated by modeling, combining with other cross-checking or platform environment data. In fact, most of the noises exhibit broad-spectrum colored noises in the amplitude spectral density of the measurement frequency band. Therefore, from the perspective of data analysis, effective suppression of unmodelable broad-spectrum random noises through learning from the noise and further improving the detection signal-to-noise ratio will have significant practical implications for the application of high-precision spaceborne electrostatically suspended inertial sensors.


\section{Methodology}
This section covers the following topics: Firstly, we introduce an overall noise reduction framework,including the theoretical foundation of the Noise2Noise algorithm and the sub-sampler. Then we introduce the U-net model and the CAE model with added reconstruction layers separately.

The overall framework of the algorithm is shown in Figure 2. The accelerometer measurement signals are divided into periodic signals and general signals, and the training set is constructed using a periodic sub-sampler and an odd-even sub-sampler, respectively. An appropriate network model is trained through the training set (adding reconstruction layers to the network model for periodic signals, and not adding them for general signals). The accelerometer measurement signals are then inputted into the trained network model to obtain the noise-suppressed signals.
\begin{figure}[htbp]
\includegraphics[width=10.5 cm]{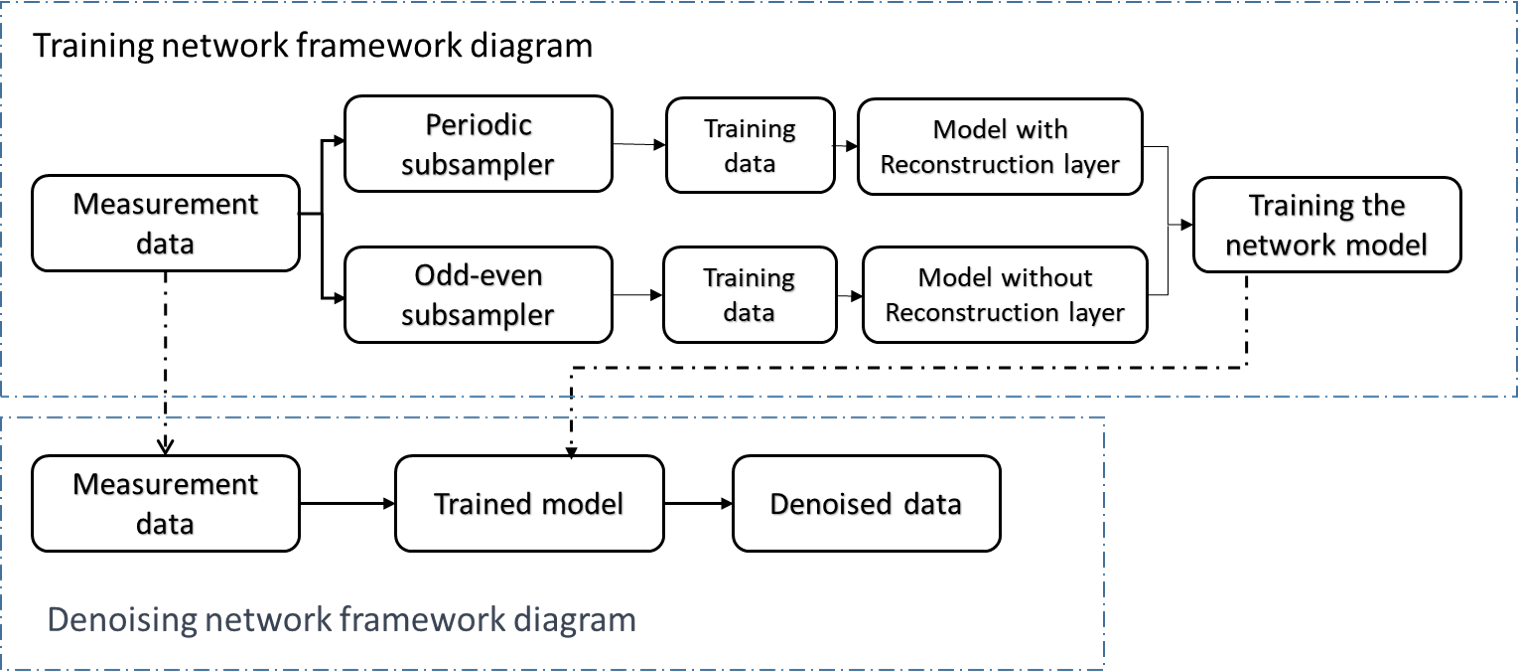}
\caption{Overall framework of noise reduction algorithm.}
\end{figure}
\subsection{Noise2Noise revisit}

Based on the content of Section \uppercase\expandafter{\romannumeral2}, the working mode of the inertial sensor in accelerometer mode for on-orbit measurement data can be expressed as follows:
\begin{equation}
y_i = x_i + z_i,
\end{equation}
where $x_i$ represents the actual non-gravitational disturbance acceleration signal $a_{para,SC}$ experienced by the spacecraft, $z_i$ represents the accelerometer noise acceleration signal $a_{para,TM}$ due to test mass, and $y_i$ represents the measured accelerometer signal $a_c$. The subscript i denotes the data label $(i=1,2,3...k)$. The basic idea of deep learning method to deal with noise is to establish the relationship between $y$ and $x$, and finally the denoised accelerometer signal can be obtained through this relationship.

The supervised learning method uses clean signal and noisy signal pairs $(x_i,x_i+z_i)$ to train the parameters of the network model, also known as Noise2Clean(N2C) method, and its specific expression is as follows:

\begin{equation}
\underset{\theta}{\operatorname{argmin}} \sum_{i=1}^m L\left(f_\theta\left(x_i+z_i\right), x_i\right),
\end{equation}
where $L$ represents a measure of the difference between the output of the network model and the true signal, $m$ represents the number of training samples , $f_\theta$ represents the signal denoising network parameterized by $\theta$. We can choose several $p$-norms as the loss function $L$. N2N model is an unsupervised deep learning method, which does not require clean signals, while requires only the noisy measurements $x_i+z_{i1}$ and $x_i+z_{i2}$ based on the true value $x^i$ to train the network model. Its specific expression is as follows:
\begin{equation}
\underset{\theta}{\operatorname{argmin}} \sum_{i=1}^m L\left(f_\theta\left(x_i+z_{i1}\right),       (x_i+z_{i2})\right).
\end{equation}

Equation (3) is equivalent to Equation (4) when the two following conditions are satisfied:
\begin{itemize}
    \item \textbf{Condition 1}:The noise of measurement in the input is independent from the noise in the target which is used to train the network;
    \item \textbf{Condition 2}:The expectation of noise added to the signal is zero.
\end{itemize}
N2N model can learn to remove noise from a signal rather than learn from noise $x+z_{1}$ to noise $x+z_{2}$ mapping. For one pair of noise samples $(y_1,y_2)$, in order to avoid the problem of insufficient samples, we can generate two pairs of noise samples $(y_1,y_2)$ and $(y_2,y_1)$ by using noise $y_1$ and noise $y_2$ as input and target respectively \cite{calvarons2021improved}:
\begin{equation}
\underset{\theta}{\operatorname{argmin}} \sum_{i=1}^m  \left( \frac{1}{2} L\left(f_\theta\left(x_i+z_{i1}\right),           (x_i+z_{i2})\right) + \frac{1}{2} L\left(f_\theta\left(x_i+z_{i2}\right),           (x_i+z_{i1})\right) \right).
\end{equation}

The N2N method requires multiple noise samples, and we use a periodic signal sub-sampler to achieve the generation of noise sample pairs from a single noise sample with periodicity. We define a periodic noise sequence as $y$ and its period is $T$, and use a hyper-parameter $k$ to control the number of periods of the interval, $k \geq 2$. The input and output of the network model is $d$ dimensional time series vector. For example, if we choose $i$-th to $(i+d-1)$-th as sample $s_1(y)$, correspondingly, if we choose $(i+ kT)$-th to $(i+ kT+d-1)$-th as sample $s_2(y)$, we get two sample pairs $(s_1(y),s_2(y))$ and $(s_2(y),s_1(y))$. Based on the sample pairs, we obtain the following equations:
\begin{equation}
f_\theta\left(x_t +z_t \right) \rightarrow x_{t+kT} + z_{t+kT},
\end{equation}
\begin{equation}
f_\theta\left(x_t +z_t \right) \rightarrow x_{t} + z_{t+kT} + (x_{t+kT}-x_t).
\end{equation}

We can ignore the difference between the truth values of the signal between intervals $n$ periods, and the map learned by the trained neural network model is $f_\theta\left(x_t +z_t \right) \rightarrow x_t + z_{t + kT}$, which satisfies the conditions of the N2N method. $nums$ is the total number of cycles. hyper-parameter $k \in [2,nums-1]$ can be selected to obtain $(2nums-4)$ samples, avoiding the problem of under-fitting of the model and poor noise reduction effect caused by insufficient samples. For a general signal without periodicity, we use odd-even samplers to generate sub-samples, dividing the noise signal $y$ into $y_{odd}$ and $y_{even}$. We can get the following equation:
\begin{equation}
f_\theta\left(x_{odd} +z_{odd} \right) \rightarrow x_{even} + z_{even},
\end{equation}
\begin{equation}
f_\theta\left(x_{odd} +z_{odd} \right) \rightarrow x_{odd} + z_{even} + (x_{even}-x_{odd}).
\end{equation}

We can assume that $(x_{even}-x_{odd})$ is approximately equal to zero, and the mapping is $f_\theta\left(x_{odd} +z_{odd} \right) \rightarrow x_{odd} + z_{even}$, which satisfies the condition of the N2N method.

\subsection{Network Model Architecture}
In the selection of deep learning models, we try the improved CAE model and U-net model used by Lehtinen in Noise2Noise, which both use one-dimensional convolution to adapt to the input and output of one-dimensional time series. In the model we try to use the dropout layer which randomly sets the weight of some neurons to 0 with probability $p$ during the training process to prevent the neural network from overfitting \cite{srivastava2014dropout}. The dimension reduction of the model uses a convolution layer with a step size of 2 instead of a pooling layer, which can avoid the loss of part of the signal information. Although using convolutional layers increases the computational complexity, it can achieve better results, and we do not have to process the data online, so it is not sensitive to the time it takes to train the network.

The difference between the two models mainly lies in the network model architecture, where the overall architecture of U-net is shown in Figure 3, which can be divided into three parts: encoder, decoder and skip connection. The input $x1$ of U-net is 1500 dimensional, representing time series waveform, and goes into the encoder. Each layer of the encoder consists of a one-dimensional convolution layer, dropout layer and pooling layer (or a convolution layer and the stride width be 2-3) to generate a feature layer, and then goes into the decoder, which consists of deconvolution and up-sampling layer. Skip connections concatenate the decoder and encoder parts of the same dimension. These connection channels allow the network to learn deep features and shallow features of the training data. Finally, the m-dimensional output $y1$ is obtained to achieve signal noise reduction.

The overall architecture of CAE is shown in Figure 3, which also has an encoder and decoder. The encoder compresses the m-dimensional input into a feature layer through a 1-D convolutional layer with a step size of 1 or 2, and then enters a decoder consisting of deconvolution proportional to the encoder. The output of the decoder additionally goes into a reconstructor composed of fully connected layers, which allows the model to converge faster under the influence of low-frequency noise.

\begin{figure}[htbp]
    \begin{minipage}[t]{0.5\linewidth}
        \centering
        \includegraphics[width=\textwidth]{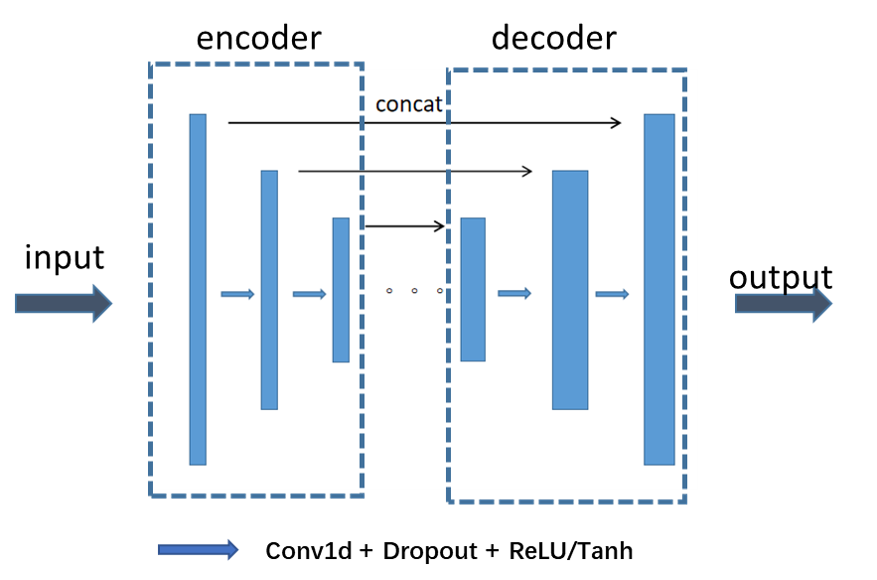}
        \centerline{(a) U-net Architecture}
    \end{minipage}%
    \begin{minipage}[t]{0.5\linewidth}
        \centering
        \includegraphics[width=\textwidth]{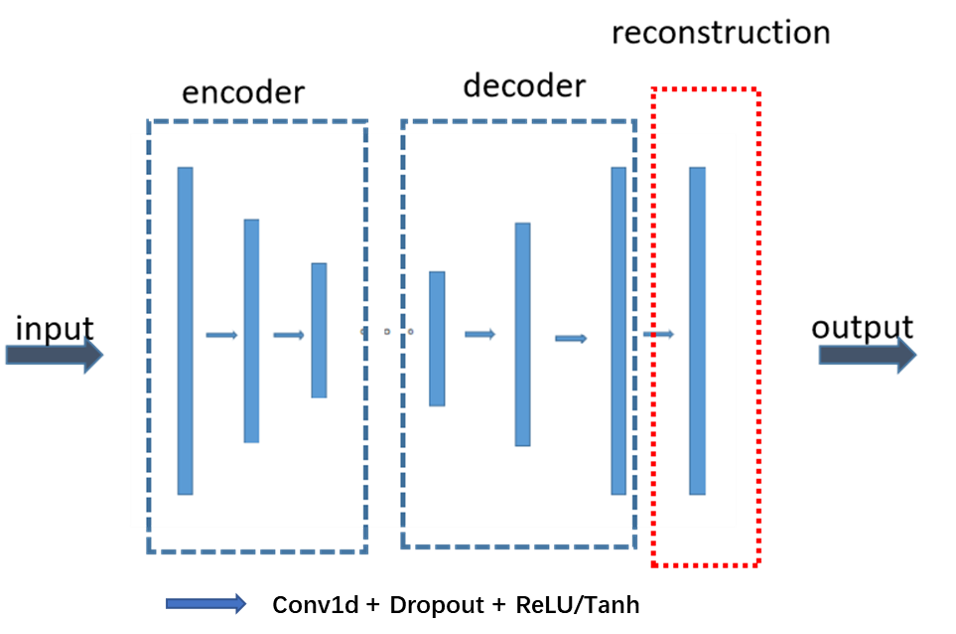}
        \centerline{(b) CAE Architecture}
    \end{minipage}
    \caption{Network model under N2N framework for signal denosing.}
\end{figure}
\section{Experiments and Results}
In the following sections, we show the effect of the N2N method for synthetic periodic noise signals and the specific noise reduction effect of real accelerometer sensor signals from Taiji-1 and GRACE-FO.
\subsection{Simulation data experiments}
In the calibration scheme of Taiji-1, we need to perform a specific maneuver scheme for the satellite. Considering the waveform of the accelerometer signal during the satellite maneuver experiment and adding more details in the period of the signal, We add the sinusoidal components $y_1$, $y_2$ and $y_3$ to form a square wave signal $y$. The synthetic noise consists of the following two parts:(1) Gaussian noise with a fixed level $\sigma=0.2$ ;(2)Colored noise generated by Gaussian noise($\sigma=0.1$) through filtering. We add synthetic noise to the square wave signal and the waveform diagram is shown in Figure 4:
\begin{equation}
    y_1 = 4 \pi \sin(x),
\end{equation}
\begin{equation}
    y_2 = \frac{4}{3} \pi \sin(3x),
\end{equation}
\begin{equation}
    y_3 = \frac{4}{5} \pi \sin(5x),
\end{equation}
\begin{equation}
    y = \frac{y_1+y_2+y_3}{25}.
\end{equation}

\begin{figure}[htbp]
    \begin{minipage}[t]{0.5\linewidth}
        \centering
        \includegraphics[width=\textwidth]{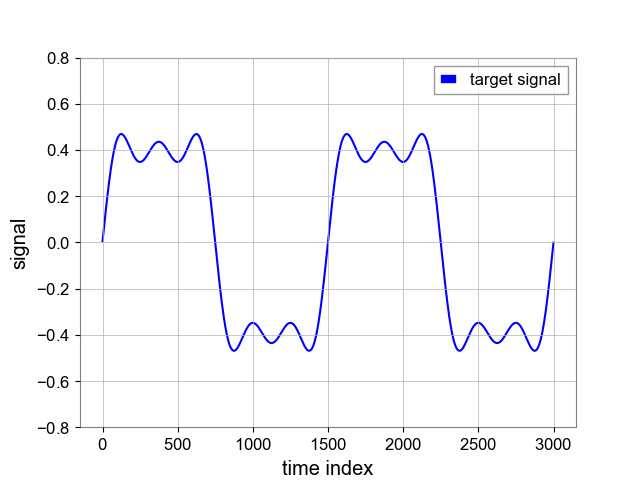}
        \centerline{(a) target signal}
    \end{minipage}%
    \begin{minipage}[t]{0.5\linewidth}
        \centering
        \includegraphics[width=\textwidth]{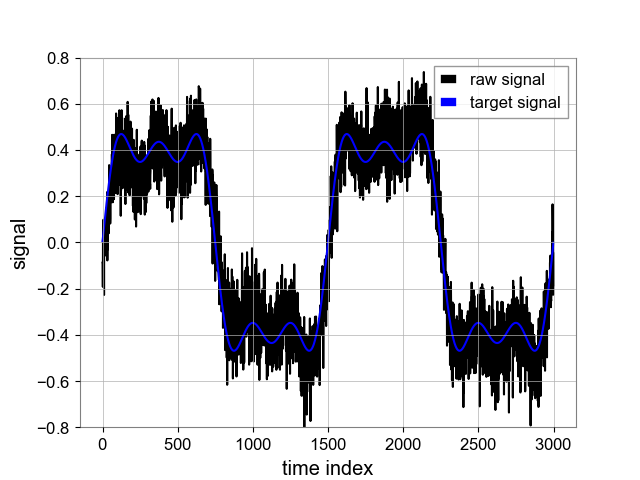}
        \centerline{(b) noisy signal}
    \end{minipage}
    \caption{Ground truth signal and mixed noise signal.}
\end{figure}

The total length of the simulation signal sequence is 150000s, the sampling frequency is 1Hz, and the signal period is 1500s. The data set is divided into a training set and a test set, which are 80\% and 20\% of the total sequence length, respectively. 80\% of the data is periodically sampled to generate a paired training set, which is used to train the neural network model while 20\% of the data is used to test the model. We choose mean square error (MSE) and signal noise ratio (SNR) to evaluate the noise reduction result of the N2N algorithm.

\begin{equation}
    MSE=\frac{1}{m} \sum_{i=1}^m\left(x_i-f\left(y_i\right)\right)^2,
\end{equation}
\begin{equation}
    SNR = 10 \log (\frac{P_s}{P_n}),
\end{equation}
where $P_s$ represents the power of the signal and $P_n$ represents the power of the noise. We also try to denoise the data by low-pass filter, wavelet decomposition denoising and Kalman filter to compare the advantages and disadvantages of N2N method and discuss whether it can be combined with N2N.

\subsubsection{Wavelet Denoising Filter}
In the process of wavelet denoising, the noise signal is decomposed by wavelet transform by selecting the appropriate wavelet, and the decomposed signal is divided into high frequency part and low frequency approximate part. Low frequency approximate part can be further decomposed, and the threshold denosing method is used to deal with the noise in high frequency part \cite{ebadi2013review}. In this study we tried different wavelet types and thresholds and finally chose "db36" wave with a threshold of 0.3. The SNR is 12.64 and MSE is 0.0078 which are achieved by the wavelet filter and the results can be seen in Figure 5.

\begin{figure}[htbp]
    \begin{minipage}[t]{0.5\linewidth}
        \centering
        \includegraphics[width=\textwidth]{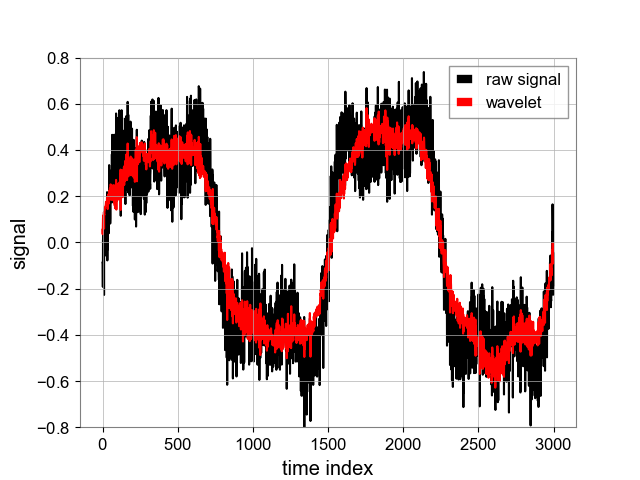}
        \centerline{(a) denoised signal}
    \end{minipage}%
    \begin{minipage}[t]{0.5\linewidth}
        \centering
        \includegraphics[width=\textwidth]{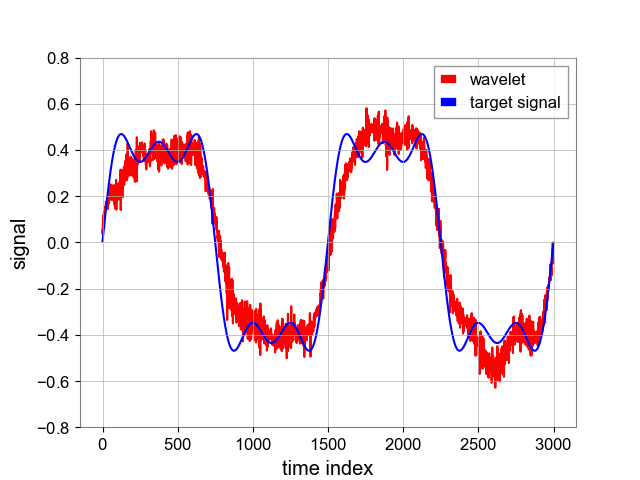}
        \centerline{(b) comparison}
    \end{minipage}
    \caption{Denoised signal using wavelet transform and comparison between
denoised signal (Red) and target signal(Blue).}
\end{figure}

\subsubsection{Kalman Filter}
The Kalman filter is divided into the prediction process and measurement process, which iterates continuously to obtain more accurate state estimation, and finally can effectively deal with noise. We tried different process variance matrix (Q) and measurement variance matrix (R) and we choose the value $Q=10$ and $R=1000$ finally \cite{alfian2021noise}. The SNR is 16.54 and MSE is 0.0033 which are achieved by the wavelet filter and the results can be seen in Figure 6.

\begin{figure}[htbp]
    \begin{minipage}[t]{0.5\linewidth}
        \centering
        \includegraphics[width=\textwidth]{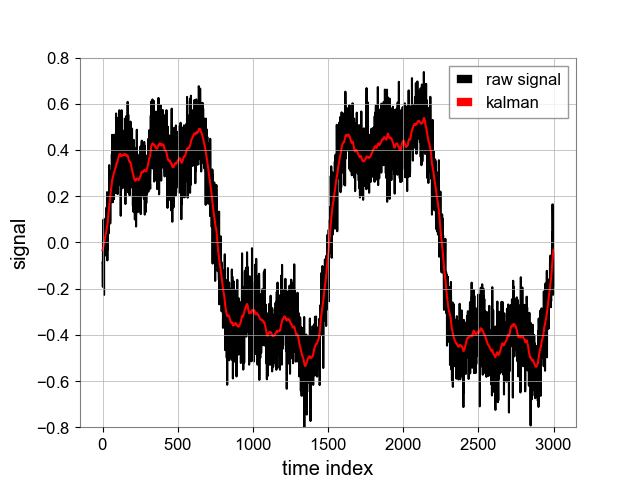}
        \centerline{(a) denoised signal}
    \end{minipage}%
    \begin{minipage}[t]{0.5\linewidth}
        \centering
        \includegraphics[width=\textwidth]{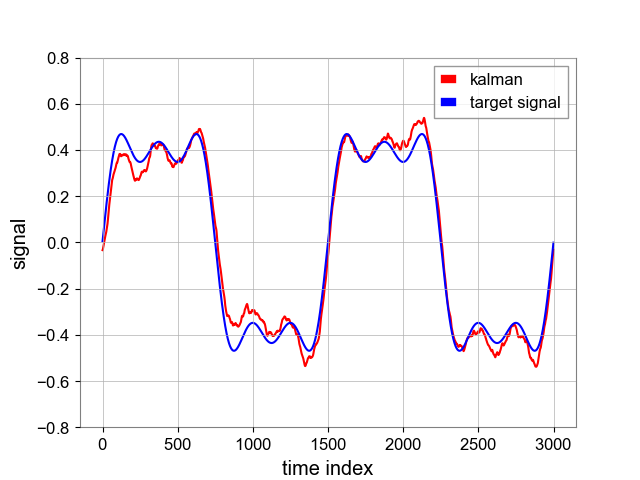}
        \centerline{(b) comparison}
    \end{minipage}
    \caption{Denoised signal using kalman filter and comparison between
denoised signal (Red) and target signal(Blue).}
\end{figure}

\subsubsection{Butterworth Filter}
The Butterworth filter makes the signal frequency flat after the passband, while the signal between the cutoff frequencies is rejected. For the signal of the experiment. For the mixed noise,  We choose a low-pass filter with a cutoff frequency of 0.005Hz. The SNR is 16.58 and MSE is 0.0033 which are achieved by the wavelet filter and the results can be seen in Figure 7.

\begin{figure}[htbp]
    \begin{minipage}[t]{0.5\linewidth}
        \centering
        \includegraphics[width=\textwidth]{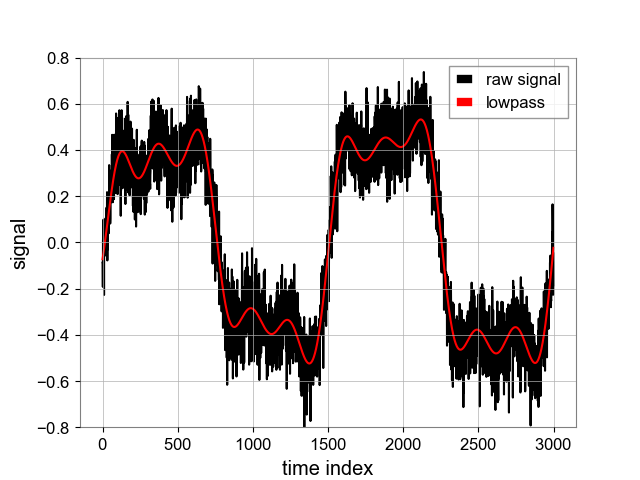}
        \centerline{(a) denoised signal}
    \end{minipage}%
    \begin{minipage}[t]{0.5\linewidth}
        \centering
        \includegraphics[width=\textwidth]{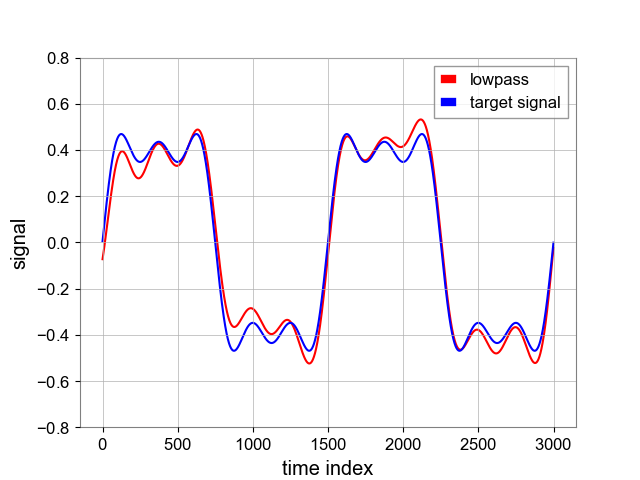}
        \centerline{(b) comparison}
    \end{minipage}
    \caption{Denoised signal using butterworth filter and comparison between
denoised signal (Red) and target signal(Blue).}
\end{figure}

\subsubsection{N2N algorithm}
All the above experiments are run on NVDIA RTX2060 GPU, the running environment of the program is Python3 under Window system, and libraries such as TensorFlow and Numpy are used. During the process of network training, we utilized the Adam optimizer \cite{kingma2015adam} with a learning rate of 0.0005 and a batch size of 16, resulting in improved performance. The choice of epoch is of great importance. When the number of epochs is large, we find that the loss value of the model loss function is very low and the convergence of the model is good, but the ultimate noise reduction effect becomes worse. Taking MSE as an example, the MSE of the test set will be lower than the MSE of the noise signal and the true value, because the MSE of the test set is obtained by calculating the input and output noise sequence signal, and the model learns from one noise distribution to another noise distribution, and loses the effect of noise reduction. Therefore, we use the early stopping mechanism, the epoch is 25, then overfitting phenomenon is avoided, the model can learn the noise reduction ability, the performance of N2N algorithm is shown in Figure 8 \cite{duvenaud2016early}. The SNR is 23.56 and MSE is 0.0006 which are achieved by the N2N algorithm with CAE model while the SNR is 17.62 and MSE is 0.0024 based on U-net model. Unlike the CAE model, the U-net model has a similar noise reduction effect as low-pass filtering.

\begin{figure}[htbp]
    \begin{minipage}[t]{0.5\linewidth}
        \centering
        \includegraphics[width=\textwidth]{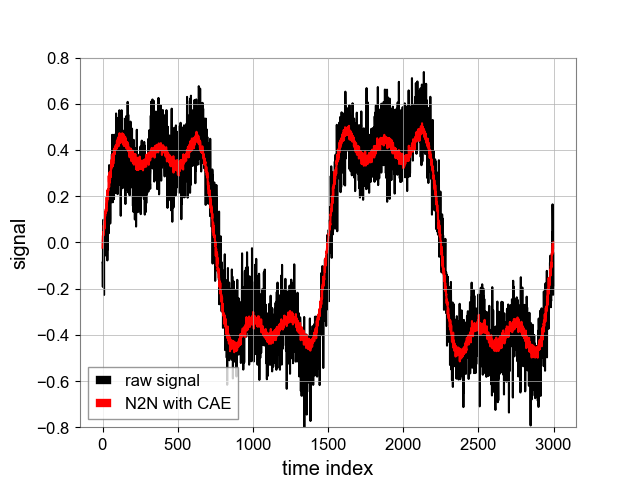}
        \centerline{(a) denoised signal}
    \end{minipage}%
    \begin{minipage}[t]{0.5\linewidth}
        \centering
        \includegraphics[width=\textwidth]{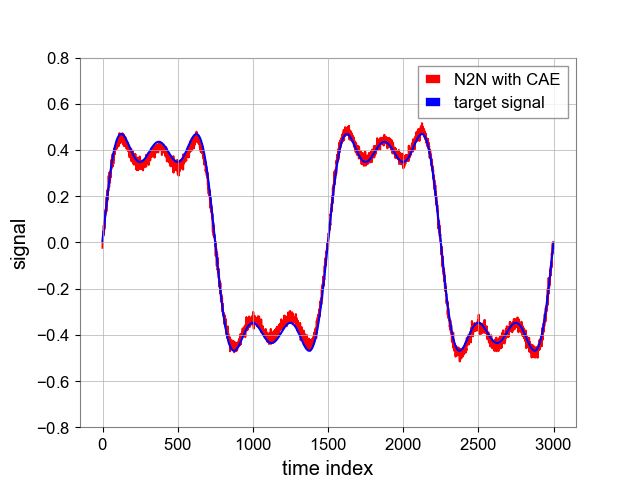}
        \centerline{(b) comparison}
    \end{minipage}
    \caption{Denoised signal using CAE and comparison between
denoised signal (Red) and target signal(Blue).}
\end{figure}

\begin{figure}[htbp]
    \begin{minipage}[t]{0.5\linewidth}
        \centering
        \includegraphics[width=\textwidth]{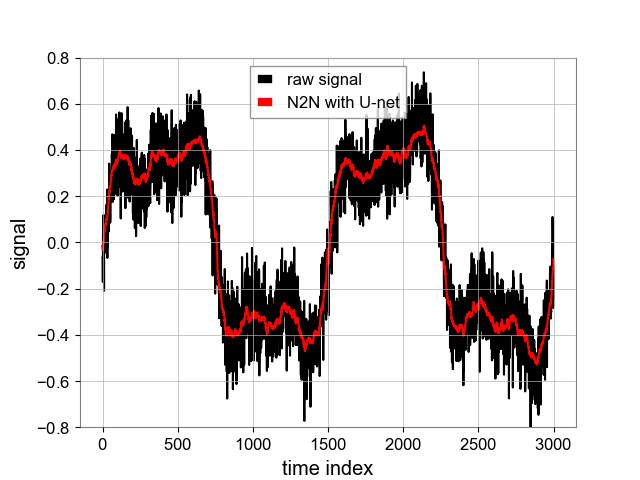}
        \centerline{(a) denoised signal}
    \end{minipage}%
    \begin{minipage}[t]{0.5\linewidth}
        \centering
        \includegraphics[width=\textwidth]{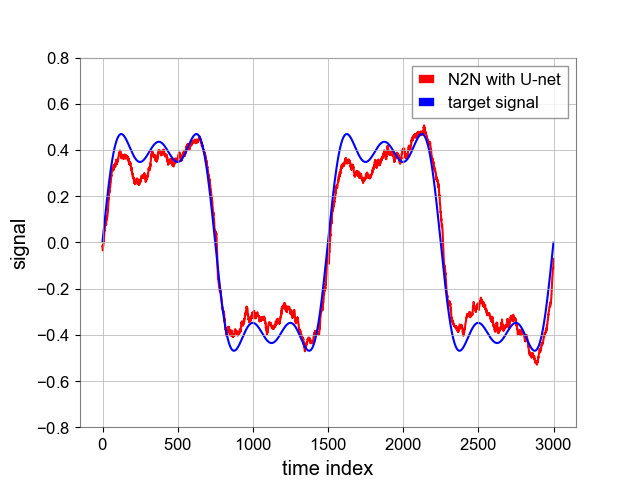}
        \centerline{(b) comparison}
    \end{minipage}
    \caption{Denoised signal using U-net and comparison between
denoised signal (Red) and target signal(Blue).}
\end{figure}

\subsubsection{Comparison between Filters and N2N}
Since the reconstruction layer of the model introduces high-frequency noise, we try to combine N2N and low-pass filter. It can be observed from Table 1 that N2N algorithm perform better than other filters in terms of noise reduction.


\begin{table}[htbp]
\begin{center}
\caption{Comparison between Filters and N2N.}
\label{tab:IS}
\begin{tabular}{cccccc}
\hline
\textbf{Filters}	& \textbf{Wavelet Transform}	& \textbf{Kalman Filter}     & \textbf{Butterworth Filter}   &\textbf{N2N} &\textbf{N2N+lowpass filter}\\
\hline
SNR		& 12.64		& 16.54		& 16.58    & 23.56  & 25.28\\
\hline
MSE		& 0.0078	& 0.0033	& 0.0033    & 0.0006   &0.0004\\
\hline

\end{tabular}
\end{center}
\end{table}

\subsection{Real data experiments}
\subsubsection{Taiji-1 data}
We take the Taiji-1 satellite as the experimental platform and upload maneuver commands to it, on May, 18th, 2022. Based on the data from the readout systems of AOS(attitude and orbit control subsystem) and inertial sensor in Taiji-1 satellite and the corresponding algorithm, where we can obtain the acceleration of TM and angular velocity of satellite platform, we can calibrate the deviation between the COM of spacecraft and Test Mass \cite{zhang2023systematic}.

Firstly, we choose the maneuver section which is periodic and highly noisy. Secondly, the data has 7 cycles and there are obvious outliers in the beginning segment of each cycle. We remove them and fill them with non-null previous values. Then, the data is scaled to $(-1, 1)$ to fit the input and output size of the neural network model, which has 1500 dimensions. Finally for the parameters of the network model, the 12 sets of data make it difficult for the neural network to converge, we take such a way of data augmentation: when using the periodic sampler, we can choose to generate a pair of training samples at an interval of n periods, where $k= 2,3,4 \cdots $. In this way we generated 44 samples. We choose the CAE model mentioned above, and the specific training details are as follows, similar to the simulation experiment, early stopping mechanism is adopted, epoch is 25, dropout is 0.1, and batchsize is 16. We first analyze the acceleration data from the Y-axis of the inertial sensor and the specific noise reduction results are as follows:

\begin{figure}[htbp]
    \begin{minipage}[t]{0.5\linewidth}
        \centering
        \includegraphics[width=\textwidth]{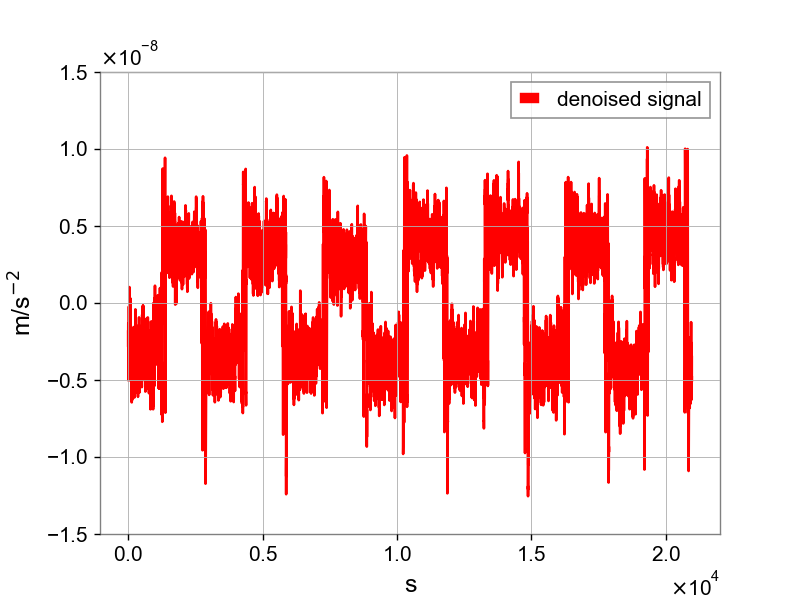}
        \centerline{(a) denoised acceleration signal}
    \end{minipage}%
    \begin{minipage}[t]{0.5\linewidth}
        \centering
        \includegraphics[width=\textwidth]{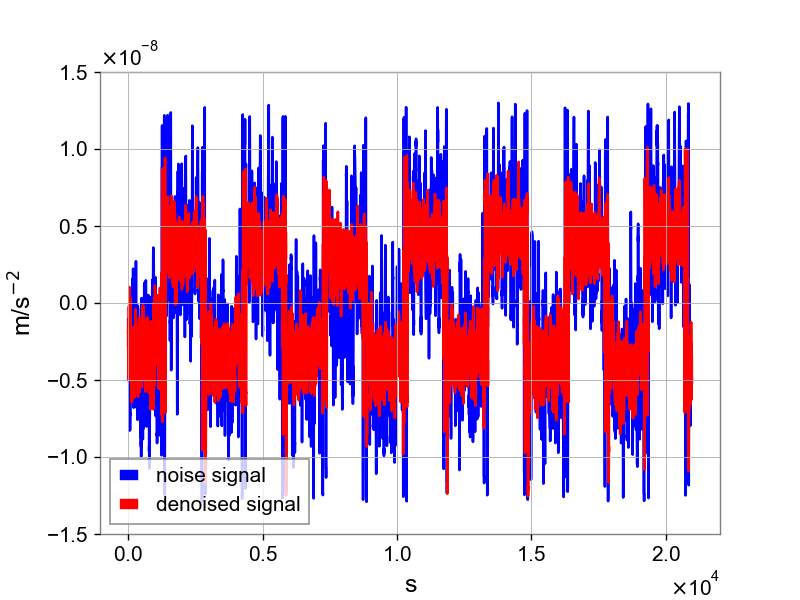}
        \centerline{(b) compared acceleration signal}
    \end{minipage}
    \caption{Measured accelerations of the inertial sensor's y-axis with N2N and without N2N. }
\end{figure}

As shown in Figure 10, the N2N method can suppress the noise in the low frequency band and improve the signal-to-noise ratio of the target signal. We can use low-pass filter and CRN filter to remove high-frequency noise which has little impact on target signal noise. After processing, the periodicity of data seems better. In the calibration scheme for COM calibration, we need the angular velocity measured by star tracker and the linear acceleration measured by the capacitive sensor in inertial sensor. Before calibration, the N2N algorithm was used for data processing and the periodic types of data are square and triangular wave signals.

\begin{figure}[htbp]
    \begin{minipage}[t]{0.5\linewidth}
        \centering
        \includegraphics[width=\textwidth]{Definitions/new-picture/taiji-time-1.png}
        \centerline{(a) square wave signal}
    \end{minipage}%
    \begin{minipage}[t]{0.5\linewidth}
        \centering
        \includegraphics[width=\textwidth]{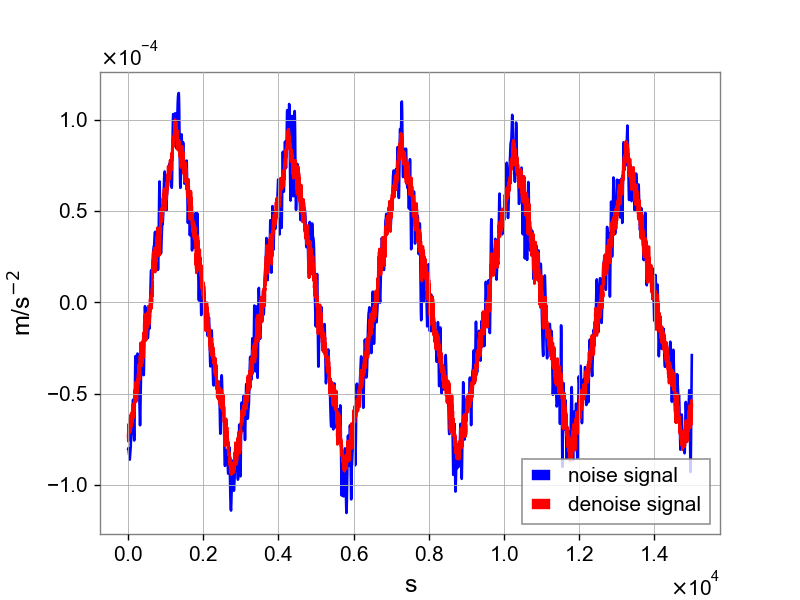}
        \centerline{(b) triangle wave signal}
    \end{minipage}
    \caption{Different periodic signal processed by N2N algorithm.}
\end{figure}

In Figure 11, there are different period types, but N2N all achieve similar results, effectively suppressing low-frequency noise and making the data periodicity better and signal-to-noise ratio higher. As shown in Figure 12 and Figure 13, we will show the results of COM calibration with and without N2N algorithm.

\begin{figure}[htbp]
\centering
\includegraphics[width=13.5 cm]{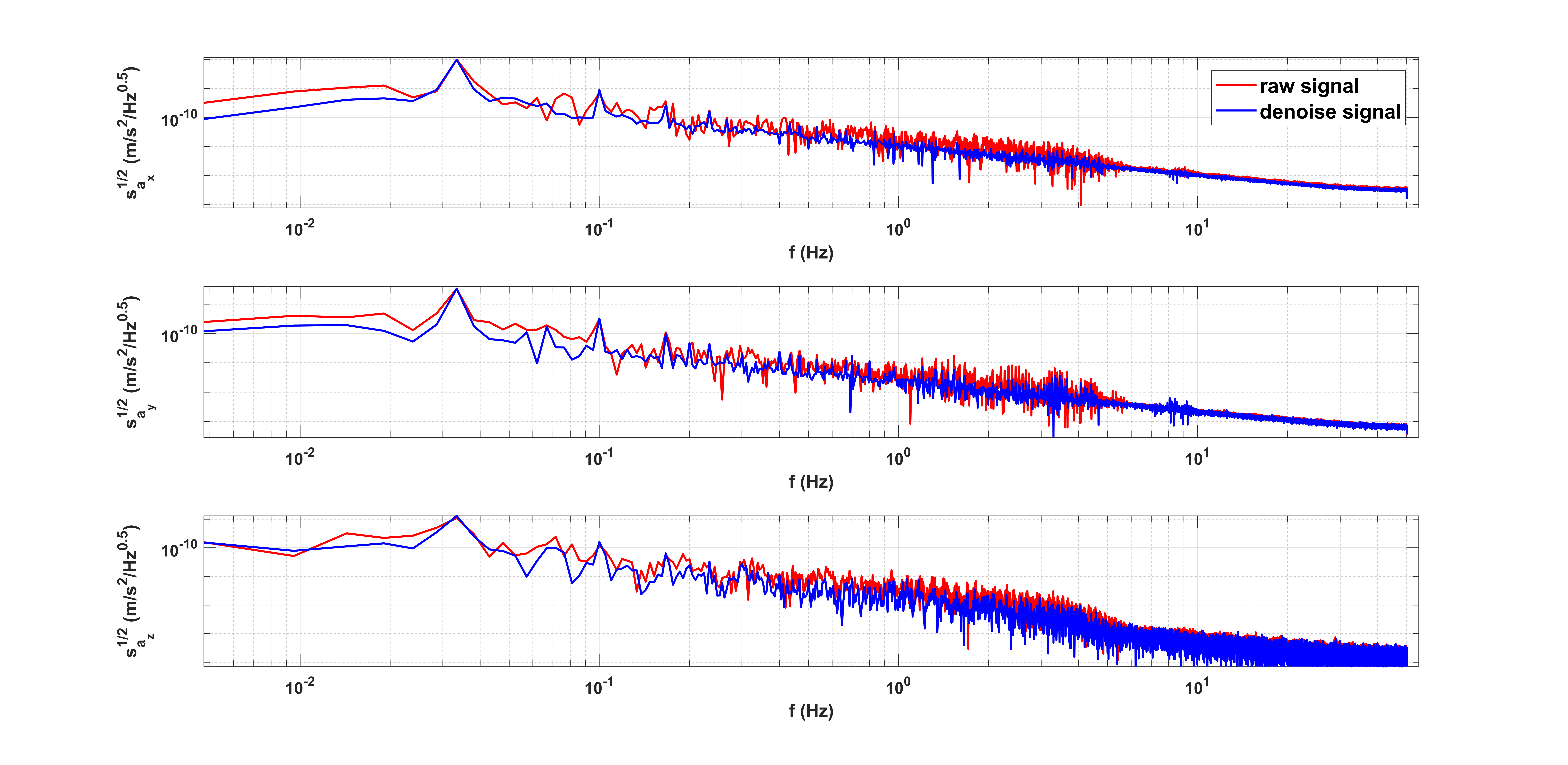}
\caption{The comparison of ASD for inertial sensor acceleration with and without N2N algorithm.\label{fig2}}
\end{figure}  

\begin{figure}[htbp]
\centering
\includegraphics[width=13.5 cm]{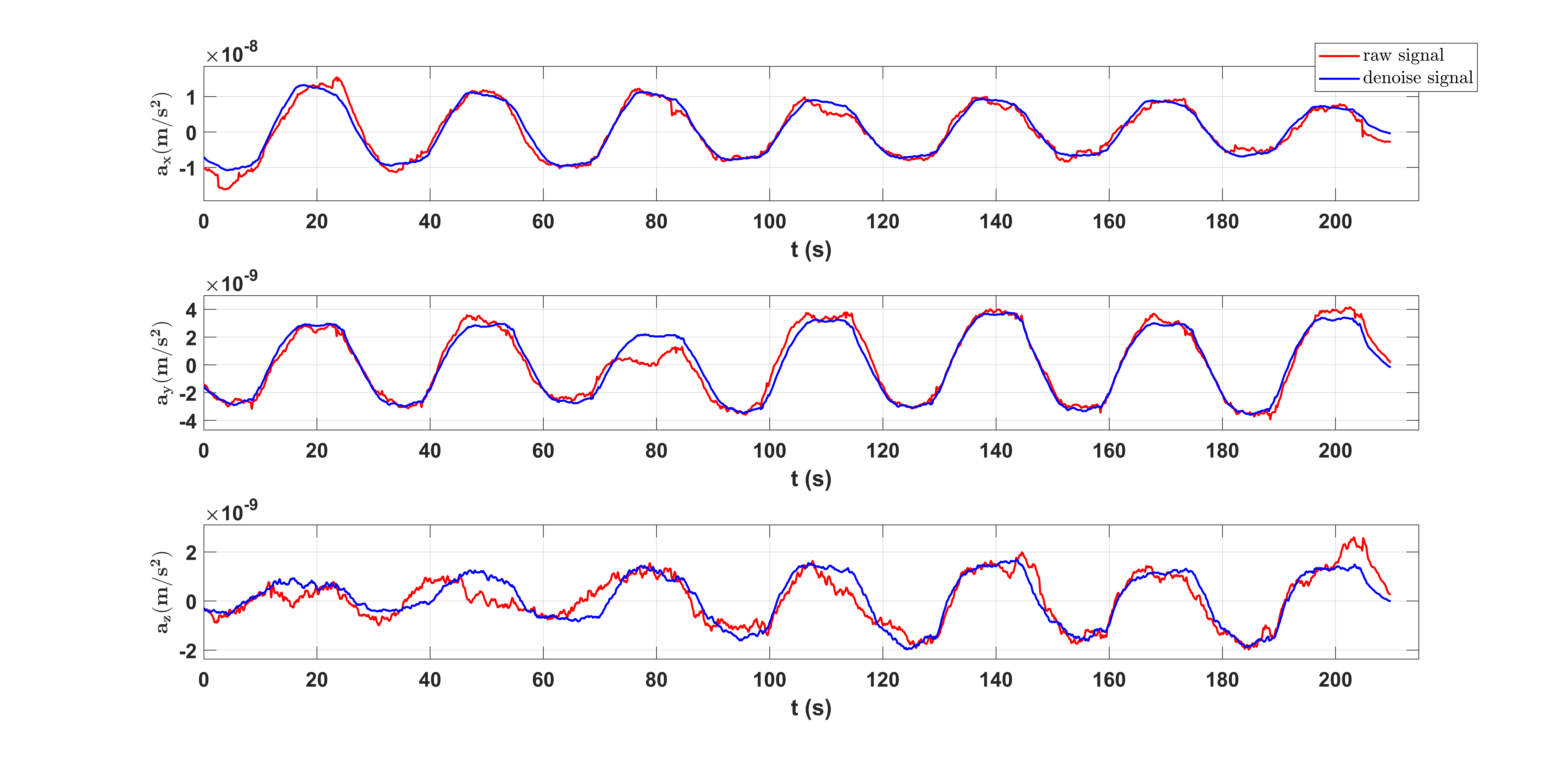}
\caption{The comparison of raw signal and denoised signal.\label{fig2}}
\end{figure}  

In Figure 12, the N2N method can better retain the peak value, effectively suppress the data on both sides of the peak, and improve the signal-to-noise ratio. At the same time, we can find that the resonance peak identification becomes clearer and the resonance peak with larger uncertainty becomes sharper and more obvious after processing. In Figure 13, we can find that after using the N2N algorithm, the data for centroid calibration task is better. Finally, the calibrated COM offset values with and without N2N are shown in Table 2.


\begin{table}[htbp]
\begin{center}
\caption{COM offset calibration results for Taiji-1 inertial sensor system with and without N2N.}
\label{tab:IS}
\begin{tabular}{ccc}
\hline
\textbf{COM offset}	& \textbf{Calibrated value with N2N(mm)}	& \textbf{Calibrated value without N2N(mm)}\\
\hline
x-axis		& -0.0793			& -0.1400\\
\hline
y-axis		& 0.3707			& 0.6270\\
\hline
z-axis       & -0.8343			& -0.8520\\
\hline

\end{tabular}
\end{center}
\end{table}

\subsubsection{GRACE-FO data}

We consider the accelerometer data from GRACE-FO's data set to analyze the noise reduction effect of the N2N method for general acceleration signals. The sampling frequency of Level-1A acceleration data is 10Hz, and we choose 5 days of data for experiment. The length of data per day is 864000, the data is limited to (-1,1) after preprocessing. By observing the data, it can be found that there are large high-frequency noise and obvious impulse noise in the signal. We did not remove the impulse noise in the preprocessing process in order to observe the denoising effect of N2N method. In the specific experiment setup, there are some differences from the above text. We generate training samples by sampling noise signals. Our sampler principle is as follows: every ten points are divided into two parts, each part is sampled once and added to the input part and label part of the dataset, and finally 80 sets of data can be obtained. The input and output dimensions of the network models are 4800 and for the CAE model we have removed the final reconstruction layer. 
The time sequence diagram and ASD diagram of accelerometer signal after noise reduction are shown as follows:

\begin{figure}[htbp]
    \begin{minipage}[t]{0.5\linewidth}
        \centering
        \includegraphics[width=\textwidth]{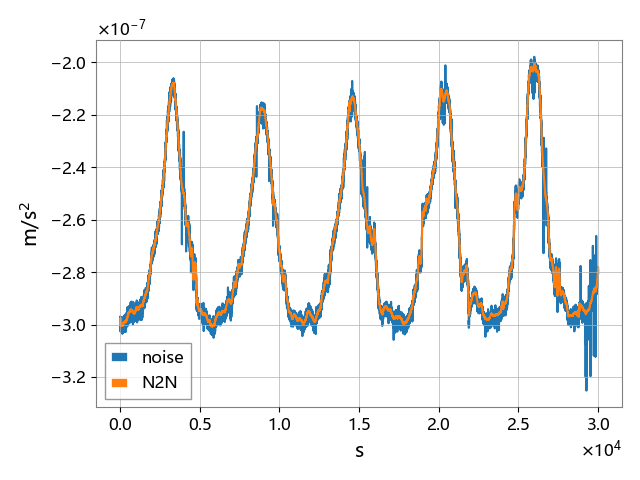}
        \centerline{(a) Comparison of acceleometer signal}
    \end{minipage}%
    \begin{minipage}[t]{0.5\linewidth}
        \centering
        \includegraphics[width=\textwidth]{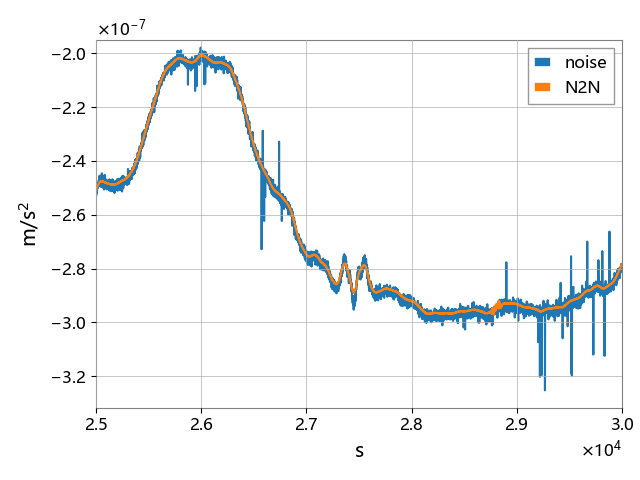}
        \centerline{(b) Comparison of local impulse noise}
    \end{minipage}
    \caption{
    Comparison of accelerometer signals before and after N2N noise reduction.}
\end{figure}

As shown in Figure 14, the N2N method is effective in dealing with impulsive and high-frequency noise, even if the noise does not fully satisfy the zero-mean condition. The analysis of the ASD spectrum shows that the accelerometer signal processed by the N2N method (yellow line) can significantly suppress noise in the frequency range above 1mHz. Compared with traditional methods such as low-pass filtering, the N2N method can preserve the high-frequency part of the signal that may exist, rather than completely filtering it out. It is observed in Figure 15 and Figure 16 that there may be variable peaks that are recognized as derived features by deep learning. The CAE model has a better noise suppression effect than the U-net model.

\begin{figure}[htbp]
\includegraphics[width=10.5 cm]{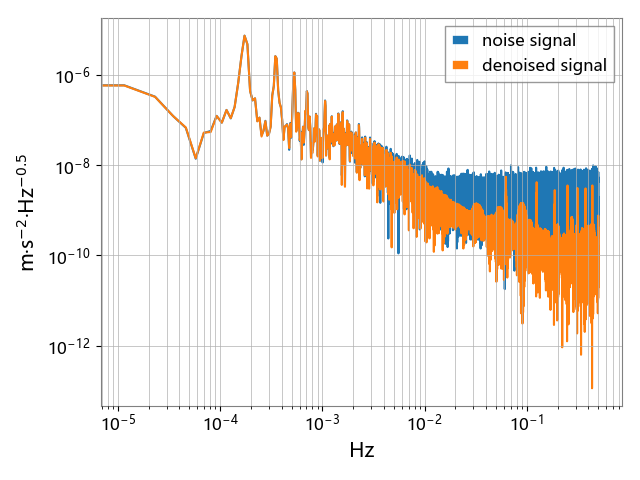}
\caption{Comparison of ASD using the CAE model.}
\end{figure}

\begin{figure}[htbp]
\includegraphics[width=10.5 cm]{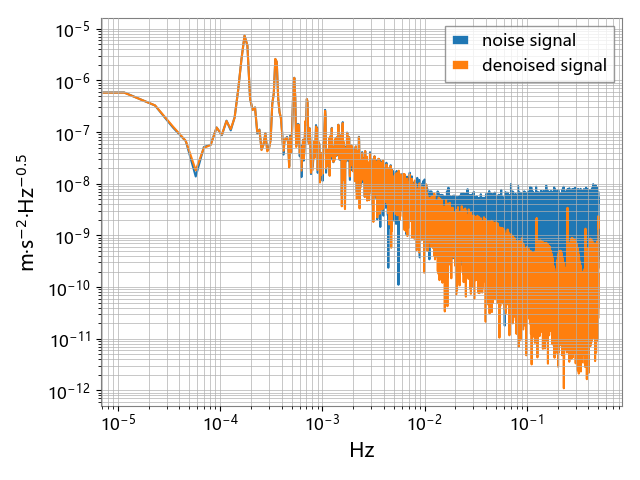}
\caption{Comparison of ASD using the U-net model.}
\end{figure}
as shown in Figure 17, the difference between the data processed by the N2N method and the Level-1B data of GRACE-FO was analyzed. Due to the loss of edge information by the convolutional layer, the residual in the middle part is smaller. To solve this problem, a sliding window approach was employed with 50$\%$ overlap between adjacent samples, so that the edge parts can be excluded. The magnitude of the residual in the middle part is at the level of $10^{-10}$.
\begin{figure}[htbp]
\includegraphics[width=10.5 cm]{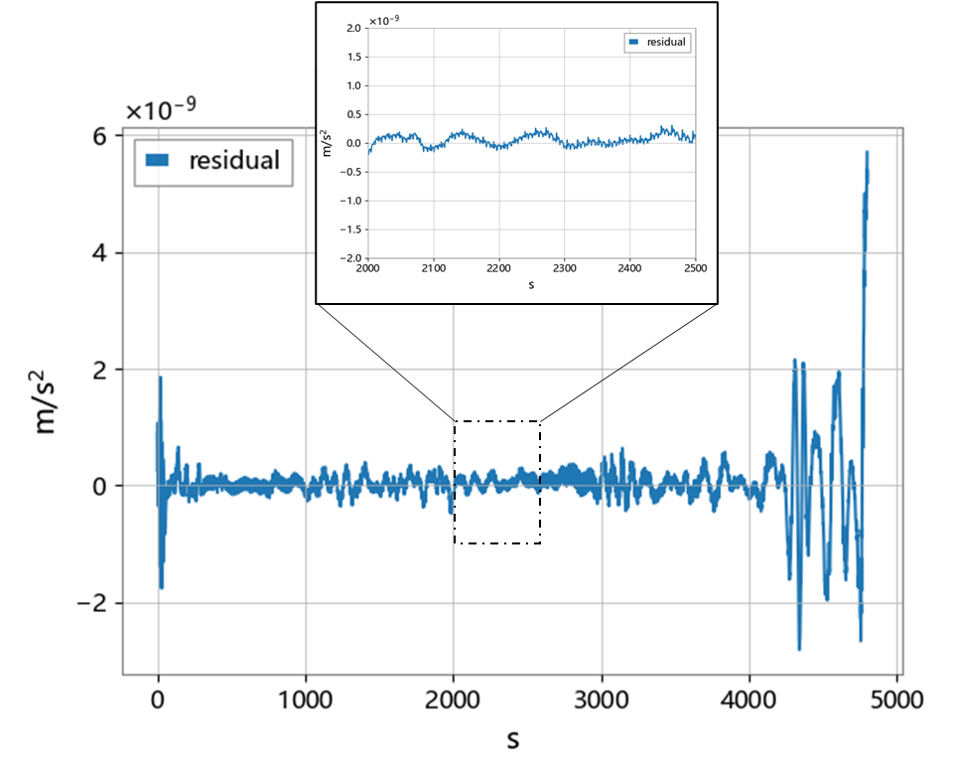}
\caption{Residual between N2N method and Level-1B.}
\end{figure}

\section{Conclusion}

Due to the complex noise mechanism of inertial sensors, it is difficult to effectively suppress noise using traditional methods. Considering the characteristics of unknown true values and single signals in on-orbit measurement signals of inertial sensors, this paper proposes a broad-spectrum noise suppression method based on N2N unsupervised learning. The measurement signal is divided into periodic signals and general signals, and different sub-sequences and network model structures are obtained using different samplers. Using a simulated data set as an example, better results were obtained compared to traditional signal processing methods, which can effectively remove the influence of mixed noise, and preliminarily demonstrate the feasibility of applying the N2N method to high-precision inertial data analysis and processing. Real data experiments were conducted on Taiji-1 and GRACE-FO, where the noise in Taiji-1 data was suppressed, and the high-frequency noise in GRACE-FO data was effectively suppressed, with data residuals at the $10^{-10}$ level compared to Level-1B data. In the future, we will study more complete N2N processing schemes and pipelines for the data processing needs of inertial sensors in gravity satellite and gravitational wave detection missions. Meanwhile, we will consider applying the N2N method more widely to high-precision sensor noise suppression.

\begin{acknowledgments}
This work is supported by the National Key Research and Development Program of China No. 2020YFC2200601, No.  2020YFC2200602 and No. 2021YFC2201901, the Strategic Priority Research Program of the Chinese Academy of Sciences Grant No. XDA15020700. The authors also acknowledge for the data resources from ``National Space Science Data Center, National Science \& Technology Infrastructure of China.''
\end{acknowledgments}




\bibliographystyle{apsrev4-1}
\bibliography{ref.bib}
\end{document}